\documentclass[floatfix, preprint, onecolumn, prd,showpacs, showkeys, preprintnumbers, nofootinbib, superscriptaddress]{revtex4-1}
\usepackage{times}
\usepackage{amssymb,amsbsy,amsmath,amsfonts}
\usepackage{graphicx}
\usepackage{float}
\usepackage{color}
\usepackage{morefloats}
\usepackage{rotating}
\usepackage{srcltx}
\usepackage{slashed}
\usepackage{multirow}
\usepackage{verbatim}
\usepackage{hyperref}
\usepackage{tabularx}

\usepackage{ulem} 

\newcommand{\quince}{{\bf 15}}
\newcommand{\dieciseis}{{\bf 16}}
\newcommand{\veinte}{{\bf 20}}

\newcommand{\sesentaytres}{{\bf 63}}
\newcommand{\cientoveinte}{{\bf 120}}
\newcommand{\cientosesentayocho}{{\bf 168}}

\newcommand{\dosmilquinientosveinte}{{\bf 2520}}
\newcommand{\cuatromilsetecientoscicuentaydos}{{\bf 4752}}

\begin{document}

\title{The $\Lambda_c(2595)$ resonance as a dynamically generated
  state: the compositeness condition and the large $N_c$ evolution}

\author{Jun-Xu~Lu}
\affiliation{School of Physics and Nuclear Energy Engineering and
International Research Center for Nuclei and Particles in the
Cosmos, Beihang University, Beijing 100191, China}

\author{Hua-Xing~Chen} 
\affiliation{School of Physics and Nuclear Energy Engineering and
International Research Center for Nuclei and Particles in the
Cosmos, Beihang University, Beijing 100191, China}
\affiliation{State Key Laboratory of Theoretical Physics, Institute
of Theoretical Physics, Chinese Academy of Sciences, Beijing 100190,
China}

\author{Zhi-Hui~Guo}
\affiliation{ Department of Physics, Hebei Normal University, Shijiazhuang 050024, China}
\affiliation{ Helmholtz-Institut f\"ur Strahlen- und Kernphysik and 
Bethe Center for Theoretical Physics,  Universit\"at Bonn, D-53115 Bonn, Germany }

\author{J.~Nieves}
\affiliation{Instituto de F\'\i sica Corpuscular (IFIC), Centro Mixto CSIC-Universidad de Valencia, Institutos de Investigaci\'on de 
Paterna, E-46071 Valencia, Spain}

\author{Ju-Jun~Xie}
\affiliation{Institute of Modern Physics, Chinese Academy of
Sciences, Lanzhou 730000, China} 
\affiliation{Research Center for
Hadron and CSR Physics, Institute of Modern Physics of CAS and
Lanzhou University, Lanzhou 730000, China} 
\affiliation{State Key
Laboratory of Theoretical Physics, Institute of Theoretical Physics,
Chinese Academy of Sciences, Beijing 100190, China}

\author{Li-Sheng~Geng} \email{lisheng.geng@buaa.edu.cn}
\affiliation{School of Physics and Nuclear Energy Engineering and
International Research Center for Nuclei and Particles in the
Cosmos, Beihang University, Beijing 100191, China}
\affiliation{State Key Laboratory of Theoretical Physics, Institute
of Theoretical Physics, Chinese Academy of Sciences, Beijing 100190,
China}
\date{\today}

\begin{abstract}
Recent studies have shown that the well established $\Lambda_c(2595)$
resonance 
contains a large meson-baryon component, which can vary depending on
the specific formalism. In this work, we examine such a picture by
utilizing the compositeness condition and the large number of colors ($N_c$)
expansion.  We examine three different  models  fulfilling
two body unitarity in coupled-channels, and  adopting  renormalization
schemes where   the mass
of the $\Lambda_c(2595)$ resonance is well described, but not
necessarily its width, since  we do not consider three body channels and work
at the isospin symmetric limit.  Both approximations might have an
effect larger on the
width than on the mass. In this context, our studies show that the compositeness of the
$\Lambda_c(2595)$ depends on the number of considered coupled channels, 
and on the particular regularization scheme adopted in the unitary
approaches and, therefore, is model dependent.  In addition, we  
perform an exploratory study of the $\Lambda_c(2595)$ in the large
$N_c$ expansion, within a scheme involving only the $\pi
\Sigma_c$ and $K\Xi'_c$ channels, whose dynamics is mostly fixed by
chiral symmetry. In this context and formulating the leading-order interaction as a
function of $N_c$, we show that for moderate $N_c> 3$ values, the mass and width of the
$\Lambda_c(2595)$ deviate from those of a genuine $qqq$ baryon,
implying the relevance of meson-baryon components in its wave
function.  Furthermore, we study the  properties of the
$\Lambda_c(2595)$, in the strict $N_c \to \infty $ limit, using an extension of the  chiral Weinberg-Tomozawa
interaction to an arbitrary number of flavors and colors. This
latter study hints at the possible existence of
a (perhaps) sub-dominant $qqq$ component in the $\Lambda_c(2595)$ resonance wave
function,  which would become dominant when the
number of colors gets sufficiently large.
\end{abstract}

\maketitle
\section{Introduction}
In the naive quark model, mesons are made up of a quark-antiquark pair
while baryons consist of three quarks. Before 2000, most hadrons could
be easily understood within such a picture, with the exception of only
a few cases, e.g., the lowest lying scalar nonet, the $\Lambda(1405)$,
and the Roper resonances~\cite{Agashe:2014kda}. The situation changed
dramatically in 2003 with the discovery of the $X(3872)$ by the BELLE
collaboration~\cite{Choi:2003ue}, that was the first of many others,
so-called $XYZ$ states, which could not be easily accommodated into
standard models of constituent quarks.  Indeed, some of them
apparently contain more than the minimum quark content dictated by the
naive quark model, such as the $Z_c(4430)$~\cite{Choi:2007wga} and
$Z_c(3900)$~\cite{Ablikim:2013mio}.  The latest $P_c$ states
discovered by the LHCb collaboration~\cite{Aaij:2015tga} are the first
exotic states of such type in the baryonic sector.  Various
theoretical interpretations of these resonances have been proposed,
ranging from weakly bound molecular or compact multi-quark states
to quark-gluon hybrids.  As many of these exotic states are located
close to the two- or even three-body strong decay thresholds,
coupled--channel effects are widely believed to play an important
role.

Unitarized approaches and their extensions, which take into account
various important constraints, such as chiral and heavy quark
symmetries, or unitarity, provide a useful framework to study
coupled--channel effects. In certain cases, the interactions among the
coupled channels can be strong enough to generate the so-called
dynamically generated states, which are customarily referred to as
molecular states as well. It is found that, somehow unexpectedly, not
only the exotic states, but also some states long believed to be
conventional hadrons, which can be explained by the constituent quark
models, turn out to contain large hadron-hadron components.  Some of
the prominent examples are the axial vector
mesons~\cite{Lutz:2003fm,Roca:2005nm,
  GarciaRecio:2010ki,Garcia-Recio:2013uva} and the low-lying tensor
states~\cite{Molina:2008jw,Geng:2008gx,
  GarciaRecio:2010ki,Garcia-Recio:2013uva}. Many studies of these
states in various decays and reactions have been performed and all the
results seem to be consistent with such a molecular picture.  

In the
heavy-flavor baryon sector, the $1/2^-$ $\Lambda_c(2595)$ and its
heavy quark symmetry (HQS) partners have been proposed to be of
molecular nature as well, although there is debate about its most
important
components~\cite{Lu:2014ina,Lutz:2003jw,Hofmann:2005sw,
GarciaRecio:2008dp,Romanets:2012hm,
  GarciaRecio:2012db, Liang:2014kra}.  More specifically in
Refs.~\cite{Lu:2014ina,Lutz:2003jw}, it is claimed that $\pi\Sigma_c$
channel plays the dominant role, while $DN$ is found to be the most
important ingredient in Ref.~\cite{Hofmann:2005sw}.~\footnote{A similar conclusion was reached in the J\"ulich meson-exchange model~\cite{Haidenbauer:2010ch}.}  After including
the $DN$ and $D^*N$ channels, as required by heavy quark spin symmetry (HQSS) arguments, the authors
of Refs.~\cite{GarciaRecio:2008dp,Romanets:2012hm,Liang:2014kra}
conclude that both of them may be needed.

In principle, wave functions are not observables themselves. As a
result, it is difficult to pin down the exact nature of a hadronic
state. The claims regarding the largest Fock components in hadron wave
functions are often model dependent. In recent years, the compositeness
condition, first proposed by Weinberg to explain the deuteron as a
neutron-proton bound state~\cite{Weinberg:1962hj,Weinberg:1965zz}, has
been advocated as a model independent way to determine the relevance
of hadron-hadron components in a molecular state. With renewed
interests in hadron spectroscopy, this method has been extended to
more deeply bound states, resonances, and higher partial
waves~\cite{Hanhart:2010wh,Baru:2003qq,Cleven:2011gp,Gamermann:2009uq,YamagataSekihara:2010pj,
  Aceti:2012dd,Xiao:2012vv,Aceti:2014ala,Aceti:2014wka,Hyodo:2011qc,Hyodo:2013nka,Sekihara:2014kya,Nagahiro:2014mba,Garcia-Recio:2015jsa}.
For the particular case of the $\Lambda_c(2595)$, the situation is a
bit unclear. For instance, it was shown in Ref.~\cite{Hyodo:2013iga}
that the $\Lambda_c(2595)$ is not predominantly a $\pi\Sigma_c$
molecular state using the effective range expansion. A similar
conclusion was reached in Ref.~\cite{Guo:2016wpy}, using a generalized
effective range expansion including Castillejo-Dalitz-Dyson pole
contributions. In this latter work, the effects of isospin breaking
corrections are also taken into account and the extended compositeness
condition for resonances developed in Ref.~\cite{Guo:2015daa} has been
applied to calculate the compositeness coefficients. Furthermore,
although in the unitary
approaches~\cite{Lu:2014ina,Lutz:2003jw,Hofmann:2005sw,GarciaRecio:2008dp,Romanets:2012hm,Liang:2014kra}
the $\Lambda_c(2595)$ is found to be of molecular nature, there is no
general agreement on its dominant meson-baryon components yet.

Another approach~\footnote{In recent years, it has been stressed that
  the quark mass dependence of a hadronic state, which can be accessed
  by present lattice QCD simulations, can also be used to distinguish
  its nature. In the present work, we are not going to approach the
  problem from  this perspective. Interested readers can see, e.g.,
  Refs.~\cite{Hanhart:2008mx,Cleven:2010aw,Altenbuchinger:2013vwa,Hyodo:2014bda,Hanhart:2014ssa,Long:2015pua}
  and references therein.}  to probe the dominant component of a
hadronic state is to study the $N_c$ dependence of the poles
associated to resonances that appear in the unitarized
meson-meson~\cite{Pelaez:2003dy,Sun:2005uk,Pelaez:2006nj,Guo:2007ff,Geng:2008ag,Nieves:2009ez,Nieves:2011gb,Ledwig:2014cla,Guo:2011pa,Guo:2015dha} 
or meson-baryon~\cite{GarciaRecio:2006wb,GarciaRecio:2006xn,Roca:2008kr,Hyodo:2007np}
scattering amplitudes, being $N_c$ the number of colors of quarks. The
$1/N_c$ expansion \cite{'tHooft:1973jz,Witten:1979kh,
  Manohar:1998xv,Jenkins:1998wy,Lucini:2012gg,Matagne:2014lla} is
valid for the whole energy region and makes specific predictions for
$q\bar{q}$  and $qqq$ states.  A genuine $q\bar{q}$ state
becomes bound as $N_c\rightarrow\infty$ with its mass scaling as
$\mathcal{O}(1)$ and its width as $\mathcal{O}(1/N_c)$. Mesonic states
of other nature may show different behavior~\cite{Jaffe:2007id}. The
mass of a generic $qqq$ state with two or three flavors evolves as
$\mathcal{O}(N_c)$ while its width scales as $\mathcal{O}(1)$ at
leading order~\cite{Witten:1979kh,Goity:2004pw,Cohen:2003fv}.

In the present work, we utilize both the compositeness condition and
the large $N_c$ behaviour to examine the nature of the
$\Lambda_c(2595)$ aiming to test the molecular scenario. This paper is
organized as follows. In Sect. II, we briefly introduce the unitarized
models used in
Refs.~\cite{Lu:2014ina,GarciaRecio:2008dp,Liang:2014kra}.  In
Sect. III, we discuss the compositeness condition and, in particular,
the effects due to the number of coupled channels considered and to the
specific regularization scheme adopted. In Sect. IV, we formulate
the large $N_c$ expansion within the model of Ref.~\cite{Lu:2014ina},
and show that in this scheme, and for a moderately large number of
colors, the $N_c$ dependence of the $\Lambda_c(2595)$ mass and width deviates from
that of a genuine $qqq$ state. We will also discuss the
$N_c\gg 3$ behavior of the $\Lambda_c(2595)$ pole position within the
dynamical model established in \cite{GarciaRecio:2008dp}, using the
findings of Refs.~\cite{GarciaRecio:2006wb,GarciaRecio:2006xn}, where
the chiral Weinberg-Tomozawa (WT) interaction is extended to an
arbitrary number of flavors and colors. This latter study {hints at the possible
existence of a (perhaps) sub-dominant $qqq$ component in the
$\Lambda_c(2595)$ resonance wave function, which would become dominant
when the number of colors gets sufficiently large.  Finally, the most
relevant conclusions of this work are collected in Sect. V.

\section{Unitarized approaches}
\label{sec:unit}
The key ingredients of unitary approaches are kernel potentials and
the  procedures adopted to restore exact two-body unitarity in coupled
channels. In practice, the kernel potentials, which
represent the strong interactions among the participating hadrons, are
generally constructed using either effective field theories, such as
chiral perturbation theory, or phenomenological Lagrangians, such as
the hidden gauge ones.  Symmetry arguments play an important role in
constructing the potentials and in fixing the unknown parameters. All
of the unitarization procedures respect coupled--channel two-body
unitarity above thresholds, but may differ in their treatment of
off-shell, left-hand cut effects, etc., which, in most cases, induce
sub-dominant corrections that  are partially accounted for by the
undetermined low energy constants. In the present work, we focus on the Bethe-Salpeter
equation method based on the so-called on-shell
approximation~\cite{Oller:1997ti, Oset:1997it,Nieves:1999bx}. For a discussion of
the off-shell effects, see, e.g., Refs.~\cite{Nieves:1998hp,Mai:2012dt} and Ref.~\cite{Altenbuchinger:2013gaa}. In the latter reference,  the off-shell effects are explicitly
demonstrated to be small.

The Bethe-Salpeter equation reads, symbolically,
\begin{equation}
T_{ij}=V_{ij}+(VGT)_{ij},
\end{equation}
where $i,j$ denote the channel index, $V$ is the kernel potential, $T$
stands for the
 unitarized amplitude, and $G$ is the two-point one-loop function.

In the study of the $\Lambda_c(2595)$, the relevant kernel potentials
$V$ have been explicitly calculated in the framework of
chiral~\cite{Lu:2014ina} and the extended hidden
gauge~\cite{Liang:2014kra} Lagrangians, and in the SU(6)$\times$HQSS
model of Ref.~\cite{GarciaRecio:2008dp}. They differ in the number of included
coupled channels and how chiral symmetry and HQSS are taken into
account. We refer to
Refs.~\cite{Lu:2014ina,Liang:2014kra,GarciaRecio:2008dp,Romanets:2012hm}
for more details. (A brief revision of the SU(6)$\times$HQSS model 
is presented in Subsect.~\ref{sec:su8nc}).

In addition to the potential, the loop function $G$ in the
Bethe-Salpeter equation also plays an important role. It has the
following simple form in 4 dimensions:
\begin{equation}
   G=i\int\frac{d^{4}q}{(2\pi)^{4}}\frac{2M}{[(P-q)^{2}-m^{2}+i\epsilon][q^{2}-M^{2}+i\epsilon]},\label{eq:loop}
\end{equation}
with $M$ and $m$ the baryon and meson masses, respectively.
This loop function is logarithmically divergent and needs to be
properly regularized. 
Two different methods can be found in the literature: 
the dimensional regularization scheme and the other in which an
ultra-violet hard cut-off is used. 
In the modified minimal subtraction scheme, the loop function  reads
\begin{eqnarray}\label{Eq:MSbar}
   \begin{split}
      G_{\overline{MS}}(s,M^{2},m^{2}) & =\frac{2 M}{16\pi^{2}}\left[\frac{m^{2}-M^{2}+s}{2s}\log\left(\frac{m^{2}}{M^{2}}\right)\right.\\
                                       & -\frac{q}{\sqrt{s}}(\log[2q\sqrt{s}+m^{2}-M^{2}-s]+\log[2q\sqrt{s}-m^{2}+M^{2}-s]\\
                                       & -\log[2q\sqrt{s}+m^{2}-M^{2}+s]-\log[2q\sqrt{s}-m^{2}+M^{2}+s])\\
                                       & \left.+\left(\log\left(\frac{M^{2}}{\mu^{2}}\right)-2\right)\right],
   \end{split}
\end{eqnarray}
where $s$ is the invariant mass squared of the meson-baryon system. To
take into account non-perturbative effects, the constant $-2$ in the above
equation is often replaced by the so-called subtraction constant $a$,
which can be slightly fine-tuned to achieve better agreement with
experimental data, in terms of  masses and widths of the dynamically
generated resonances.  An alternative way to fix $a$ is to require
that at a certain energy scale, $\mu_0^2$, the unitarized amplitude reduces to that of
the tree level, such as $G(\mu_0^2)=0$. This has been referred to as the
naturalness requirement~\cite{Hofmann:2005sw}. In the following, we refer to this
regularization method as ``DR-naturalness.'' It should be noted that
this is the method adopted in
Refs.~\cite{Hofmann:2005sw,GarciaRecio:2008dp}.

  In Ref.~\cite{Altenbuchinger:2013vwa}, a so-called HQS inspired
  regularization scheme has been suggested, which is manifestly
  consistent with 
  both the chiral power counting and heavy-quark spin-flavor (SF)
  symmetry, up to $\Lambda_{QCD}/M_H$ corrections, where $M_H$ is a generic heavy-hadron
  mass. In this scheme, referred to as ``DR-HQS'' in the present work,
  the loop function $G$ reads:
\begin{equation}
G_{HQS}=G_{\overline{MS}}-\frac{2\mathring{M}}{16\pi^{2}}\left(\log\left(\frac{\mathring{M}^{2}}{\mu^{2}}\right)-2\right)+\frac{2 m_\mathrm{sub}}{16\pi^{2}}\left(\log\left(\frac{\mathring{M}^{2}}{\mu^{2}}\right)+a\right),
\end{equation}
where $m_\mathrm{sub}$ is a generic pseudoscalar meson mass, which can
take the value of $m_\pi$ in the $u$, $d$ flavor case or an average of
the pion, kaon, and eta masses in the $u$, $d$, and $s$ three
flavor case.  $\mathring{M}$ is the chiral limit value of the charmed
or bottom baryon masses. The apparent renormalization scale dependence originates from that
  of the dimensional regularization and has little to do with  the``HQS" description (for more details, please see 
Ref.~\cite{Lu:2014ina}). Note that  in the
present case, this scheme is equivalent to the modified minimal
subtraction one discussed above. In the numerical calculations,
we use $\mathring{M}=2.5349$ GeV, which is the average of the sextet
charmed baryon masses,
$m_\mathrm{sub}=0.368$ GeV, average of the masses of the pseudoscalar mesons, and $\mu=1$ GeV.
In principle, one could use a different value for  $\mathring{M}$ in
the light baryon sector, but this would be equivalent to the use of 
different subtraction constants for different channels, which we would like to avoid.
Natural values for the
  subtraction constant, considering the range of baryon masses (i.e.,
  $\mathring{M}$)  
involved in the  present study, lie in the  $[-6,-2]$ interval, using $a=-2$ as a reference in the modified minimal subtraction scheme.

The loop function can also be regularized with an ultra-violet hard
cutoff, $\Lambda$, i.e., 
\begin{equation}
G_\mathrm{cut}=\int_0^\Lambda \frac{q^2\,dq}{2\pi^2}
\frac{E_M+E_m}{2 E_M  E_m}\frac{2 M}{s-(E_M+E_m)^2+i\epsilon}, \label{eq:cutoff}
\end{equation}
with $E_M=\sqrt{q^2+M^2}$, and $E_m=\sqrt{q^2+m^2}$.  Taking into
account the typical size of the hadrons, values around 1 GeV are
natural for $\Lambda$, although its exact value is
in most cases determined from a fit to data.
 
One of the main  objectives of this work is,  using potentials constructed in
different frameworks~\cite{Lu:2014ina,Liang:2014kra,GarciaRecio:2008dp}, to
study how the so-called compositeness or the dominance of a certain
channel varies with the scheme adopted to regularize the loop function $G$.

\begin{table}\centering
\begin{tabular}{c|c|c|c}
  \hline\hline
  Meson                  & mass(GeV)             & Baryon          & mass(GeV) \\ \hline
  $\pi$                  & $0.13804$               & $N$             & $0.93892$\\
  $K$                    & $0.495645$            & $\Lambda$       & $1.11568$ \\
  $\eta$                 & $0.54786$             & $\Sigma$        & $1.19315$\\
  $\rho$                 & $0.77549$             & $\Xi$           & $1.31829$\\
  $K^{*}$                & $0.89388$             & $\Sigma^{*}$    & $1.38280$\\
  $\omega$               & $0.78265$             & $\Xi^{*}$       & $1.53180$\\
  $\phi$                 & $1.01946$             & $\Lambda_{c}$   & $2.2865$\\
  $\eta^{'}$             & $0.95778$             & $\Xi_{c}$       & $2.46934$\\
  $D$                    & $1.86723$             & $\Sigma_{c}$    & $2.4535$\\
  $D^{*}$                & $2.00861$             & $\Sigma_{c}^{*}$& $2.51807$\\
  $D_{s}$                & $1.96830$             & $\Xi_{c}^{'}$   & $2.57675$\\
  $D_{s}^{*}$            & $2.11210$             & $\Xi_{c}^{*}$   & $2.64590$\\
  \hline\hline
\end{tabular}
\caption{Meson and baryon masses used in the present work.}\label{Table:Masses}
\end{table}

\section{The compositeness condition}

As mentioned previously, the compositeness analysis proposed by
Weinberg in Refs.~\cite{Weinberg:1962hj,Weinberg:1965zz} is only valid
for bound states. For resonances, it involves complex numbers and,
therefore, a strict probabilistic interpretation is lost.  The generalization
of the compositeness study for resonances has been put forward by
different groups.   The weight of a hadron-hadron component in a
composite particle is defined as~\cite{Aceti:2014ala}
 \begin{equation}
  X_i= \mathrm{Re}\tilde{X}_i\,,
 \end{equation}
 with 
 \begin{equation}\label{defxi}
  \tilde{X}_i=- g_i^2\left[\frac{\partial G_i^{II}(s)}{\partial \sqrt{s}}\right]_{s=s_0} \,, 
 \end{equation}
where $s_0$ is the pole position in the complex $s$ plane, $G_i^{II}$
is the loop function evaluated on the second Riemann sheet (SRS), and $g_i$
is the coupling of the  resonance to the channel
$i$, which can be obtained as
\begin{equation}
 g_i^2=\mathop{\mathrm{lim}}_{\sqrt{s}\rightarrow\sqrt{s_0}}\left(\sqrt{s}-\sqrt{s_0}\,
 \right)T_{ii}^{II}\,, 
 \end{equation}
where $T_{ii}^{II}$ is the $ii$ element of the $T$ amplitude on the
SRS. For bound states, the quantity $\tilde{X}_i$ is real and it is
related to the probability of finding the state in the channel
$i$. For resonances, $\tilde{X}_i$ is still related to the squared
wave function of the channel $i$, in a phase prescription that
automatically renders the wave function real for bound states
\cite{Aceti:2014ala}, and so it might  still be used as a measure of
the weight of that hadron-hadron channel in the composition of the
resonant state~\cite{Aceti:2014ala,Sekihara:2014kya}.

 The deviation of the sum of $X_i$ from unity is related to the energy
 dependence of the $s$-wave  potential,
 \begin{equation}
 \sum_i X_i=1-Z,  \label{eq:defXi}
 \end{equation}
 where 
 \begin{equation}
 Z={\rm Re}\tilde{Z}= {\rm Re}\left (-\sum_{ij} \left[g_i G_i^{II}(\sqrt{s})\frac{\partial
     V_{ij}(\sqrt{s})}{\partial
     \sqrt{s}}G_j^{II}(\sqrt{s})g_j\right]_{s=s_0}\right) . \label{eq:defZ}
 \end{equation}
Note that the Eqs.~(\ref{eq:defXi}) and (\ref{eq:defZ}) get support from the
sum rule~\cite{Hyodo:2011qc,Hyodo:2013nka, Sekihara:2014kya, Garcia-Recio:2015jsa}
\begin{equation}
-1= \sum_{ij} g_ig_j\left( \delta_{ij} \left[\frac{\partial
    G_i^{II}(s)}{\partial \sqrt{s}}\right]_{s=s_0} + \left[ G_i^{II}(\sqrt{s})\frac{\partial
     V_{ij}(\sqrt{s})}{\partial
     \sqrt{s}}G_j^{II}(\sqrt{s})\right]_{s=s_0}\right), \label{eq:sum-rule}
\end{equation}
which is also satisfied in the case of bound states located in the
first Riemann sheet, and guaranties that the imaginary parts of
$\sum_i \tilde{X}_i$ and $\tilde{Z}$ must cancel.  The field
renormalization constant $\tilde{Z}$ itself is well-defined even for resonances, since it corresponds to the residue of the
renormalized two-point function~\cite{Hyodo:2011qc}. Thus, there is no
fundamental problem in calculating $\tilde{Z}$ using
Eq.~(\ref{eq:defZ}), but the probabilistic interpretation of
the obtained result is not straightforward.  The field renormalization
constant $\tilde{Z}$ measures the effect of the elementary
contribution as the deviation from unity, and it is in general a
complex number. Therefore one should be aware that $\tilde{Z}$ can not
directly be interpreted as the ``probability'' of the elementary
component~\cite{Hyodo:2013nka}. Conversely, strictly speaking,
$\tilde{X}_i$ cannot be interpreted as a probability of finding a
two-body component. Nevertheless, because it represents the
contribution of the channel wave function to the total normalization,
the compositeness $\tilde{X}_i$ will have an important piece of
information on the structure of the resonance. In general, however, all $\tilde{X}_i$
and $\tilde{Z}$ can be arbitrary complex numbers constrained by
Eq.~(\ref{eq:sum-rule}). The probabilistic interpretation of the
structure of a resonance from $\tilde{X}_i$ and $\tilde{Z}$ is not
possible when the imaginary parts are sizable~\cite{Sekihara:2014kya}
or when there is a large cancellation among the real parts of
$\sum_i\tilde{X}_i$ and $\tilde{Z}$ to meet the sum rule of
Eq.~(\ref{eq:sum-rule}), but with one of them exceeding the
unity. T. Hyodo, following the ideas of T. Berggren~\cite{Berggren} in
the seventies, has proposed to look at the parameter $P$, defined as
\begin{equation}
P=|\tilde{Z}|+\Big|\sum_i\tilde{X}_i\Big|-1= |\tilde{Z}|+|1-\tilde{Z}|-1= 
\Big|1-\sum_i\tilde{X}_i\Big| + \Big|\sum_i\tilde{X}_i\Big|-1, \label{eq:defP}
\end{equation}
and try to give  a ``probabilistic'' interpretation to $\tilde{Z}$ and
$\sum_i\tilde{X}_i$ only for those cases where 
$P$ is much smaller than 1/2~\cite{Hyodo}.

 In the picture advocated in Ref.~\cite{Aceti:2014ala} imaginary parts
 are neglected. The quantity $1-Z$ is taken to represent the compositeness of the hadronic state
 in terms of all the considered channels, and $Z$ is referred to as
 its elementariness. Within this picture, a non-vanishing $Z$ takes into account that
 ultimately the model is an effective one. The energy dependent
 interaction effectively accounts for other possible interaction
 mechanisms not explicitly included in the $s-$wave hadron-hadron
 description. These could be other hadron-hadron interactions, or even
 genuine hadron components not of the molecular type
 (hence the appellative elementariness). Thus, a small value of $Z$
 indicates that the state is well described by the contributions
 explicitly considered, namely, $s-$wave hadron-hadron  channels.
 Conversely, a large value of $Z$ indicates that, for that state,
 significant pieces of information are missing in the model, and this
 information is being included through an effective interaction, to
 the extent that the experimental hadronic properties are reproduced
 by the model. However, it is not clear how to interpret $Z$ obtained
 from the smooth energy dependence of the chiral potential
 $V$~\cite{Aceti:2014wka}.  In addition, it should be emphasized that,
 for processes involving short distances, it is the wave function at
 the origin that matters ($g_i G_i$ for the $s$
 wave)~\cite{Gamermann:2009uq,Gamermann:2009fv}\footnote{For an
   extensive discussion on this issue, see Ref.~\cite{Aceti:2014wka},
   where it was concluded that to judge the relevance of each channel
   one has to study different physical processes.}.

On the other hand, in Ref.~\cite{Guo:2015daa}, it was claimed that one
can formulate a meaningful compositeness relation with only positive
coefficients thanks to a suitable unitary transformation of the $S$
matrix. This in practice amounts to take the absolute value of
$\tilde{X}_i$ in Eq.~\eqref{defxi} to quantify the probability of
finding a specific component in the wave function of a hadron.  Notice
that the recipe advocated in Ref.~\cite{Guo:2015daa} is not applicable
to all types of poles. In particular the arguments of this reference
exclude the case of virtual states or resonant signals which are an
admixture between a pole and an enhanced cusp effect by the pole
itself. More specifically, the probabilistic interpretation given in
\cite{Guo:2015daa} to
$|\tilde{X}_i|$ is only valid when $\sqrt{{\rm Re} s_0} > M_{i,{\rm
    th}}$, with $M_{i,{\rm th}}$ the corresponding threshold of the $i$th
channel~\footnote{In this situation the
  convergence region of the Laurent series of the $S$ matrix around
  the pole incorporates some intervals of the physical real axis
  around the pole mass $M_R$~($\equiv \sqrt{{\rm Re} s_0}$), and in
  these circumstances  it follows $|\tilde{X}_i| \le 1$. Actually, it can be proved
  that $\sum_{i}|\tilde{X}_i| \le 1$, where the sum is only over
  the channels fulfilling $\sqrt{{\rm Re} s_0} > M_{i,{\rm
      th}}$~\cite{Guo:2015daa}.  Thus, the so-called effective
  elementariness is then defined as $1-\sum_{i}|\tilde{X}_i|$, which
  accounts for the contributions of the heavier channels that do not
  enter into the sum. }.

In what follows, we will examine how the number of coupled
channels and the particular regularization scheme affect the predicted
(calculated) compositeness of the $\Lambda_c(2595)$. For such a
purpose, we first fix the number of coupled channels and therefore the
kernel potentials, and then compare the resulting compositeness
coefficients. The meson and baryon masses employed in the numerical
analysis  are the same as those used in
Ref.~\cite{Garcia-Recio:2015jsa} and are compiled here in Table \ref{Table:Masses}.

According to the PDG, the $\Lambda_c(2595)$ has a mass of
$2592.25\pm0.28$ MeV and a width of $2.6\pm0.6$
MeV~\cite{Agashe:2014kda}.  Therefore, the only parameter in each of
the three regularization schemes discussed in Sect.~\ref{sec:unit} is fixed in such a way that the mass
of the $\Lambda_c(2595)$ is reproduced.  We do not attempt to fix the
width because we only consider here two-body coupled channels and work
at the isospin symmetric limit, both approximations can have an
effect larger on the
width than on the mass (see an elaborate discussion in
Ref.~\cite{Guo:2016wpy}).

\subsection{Sixteen channels}
\label{sec:16}

First, we consider the sixteen channels considered in
Refs.~\cite{GarciaRecio:2008dp,Romanets:2012hm}, making also use of the 
kernel potentials provided by the
SU(6)$\times$HQSS model derived in these references, and examine the dependence
of the compositeness condition on the renormalization/regularization scheme employed
to render the loop function ultraviolet-finite. 

The SU(6)$\times$HQSS model used in
Refs.~\cite{GarciaRecio:2008dp,Romanets:2012hm} is basically a SU(8)
SF extension\footnote{This corresponds to treating the eight
  states of a quark ($u$, $d$, $s$ or $c$ with spin up, $\uparrow$ ,
  or down, $\downarrow$ ) as equivalent, and leads to the invariance
  group SU(8). Because SU(8) SF symmetry is strongly broken
  in nature, mass and weak decay constant breaking effects are taken into account
  in Refs.~\cite{GarciaRecio:2008dp,Romanets:2012hm}.} of the SU(3)
chiral WT leading order meson-baryon interaction term, including
ground state vector meson and $J^P=3/2^+$ baryon degrees of
freedom. This is actually strictly correct only when coupled channels
involving $cc\bar c$ components (e.g., doubly charmed baryons and $
{\bar D}^{(*)}$ antimesons) are neglected as done in
Refs.~\cite{GarciaRecio:2008dp,Romanets:2012hm}. These channels are OZI
disconnected from those involving just one heavy quark. Note that in
the heavy-quark limit, the OZI rule becomes exact because the number
of charm quarks and the number of charm antiquarks are separately
conserved. (For a more detailed discussion see
Ref.~\cite{Garcia-Recio:2013gaa}). In this framework, there appear
two $\Lambda_c(2595)$ states, resemblance of the two $\Lambda(1405)$
resonances found in chiral unitarity approaches, with one of them
narrower than the other~\cite{GarciaRecio:2008dp,Romanets:2012hm}. 

To
make a reliable comparison, we adjust the only parameter in each of
the regularization schemes discussed above to fix the real part of the
narrower pole to the $\Lambda_c(2595)$ resonance mass quoted in the
PDG~\cite{Agashe:2014kda}. This yields the following parameters,
$\alpha=0.97952$ for the DR-naturalness scheme\footnote{This is
  defined for instance in Eq.~(17) of Ref.~\cite{GarciaRecio:2008dp}. },
$q_\mathrm{max}=0.67898$ GeV for the cutoff scheme, and $a=-3.37865$
for the DR-HQS scheme.
 
 Compositeness results for the $\Lambda_c(2595)$ and its broader
 partner are shown in  Tables \ref{Table:narrow} and
 \ref{Table:broad}, respectively. Among the 16 coupled channels, in
 general  the most relevant ones are $\pi\Sigma_c$, $DN$ and
 $D^*N$. In the case of the narrow state (Table~\ref{Table:narrow})
 and for the DR-naturalness scheme, the first of these channels is
 suppressed, and the dominant components turn out to be  $DN$  and
 $D^*N$. 

For the sibling state of the $\Lambda_c(2595)$, it seems that the
 $\pi\Sigma_c$ channel plays the dominant role, except in the cutoff
 scheme, where it appears as a bound state and $DN$ and $D^*N$
 channels are more important. For the state that we assign to the
 $\Lambda_c(2595)$, different regularization schemes yield somehow
 different results. The $D^*N$ channel plays a leading role in the
 DR-naturalness scheme of Refs.~\cite{GarciaRecio:2008dp,
   Romanets:2012hm}. In the cutoff scheme, $\pi\Sigma_c$ is the
 dominant channel, with $D^*N$ the next component in importance. In
 the DR-HQS scheme, all three mentioned channels seem to be similarly
 important, with a large imaginary part for
 $\tilde{X}_{\pi\Sigma_c}$. On the other hand, when interpreting the
 compositeness using the prescription of Ref.~\cite{Guo:2015daa}, we
 find that the weights of $\pi\Sigma_c$ inside the $\Lambda_c(2595)$
 are 0.11, 0.71 and 0.97 for the DR-naturalness, cutoff and DR-HQS
 schemes, respectively.  Since the $DN, D^{*}N$ and other heavier
 channels do not meet the criterion of Ref.~\cite{Guo:2015daa}, no
 definite conclusions can be made separately for each of these
 channels.  Besides, $1-|\tilde{X}_{\pi\Sigma_c}|$ would be 
the effective elementariness, which get contributions from all of the other heavier
channels. Similar conclusions can be also made for the broader state
 in Table~\ref{Table:broad}.  

We pay now attention to the uncertainty
 parameter introduced in Eq.~(\ref{eq:defP}). It is significantly
 smaller than 1/2, which allows for an approximate ``probabilistic''
 interpretation of $X_i$ and $Z$ as advocated in
 Ref.~\cite{Aceti:2014ala}, only in the DR-naturalness and cutoff
 schemes for the $\Lambda_c(2595)$ and its broader partner,
 respectively. With larger uncertainties, the DR-HQS scheme for both
 resonances and the DR-naturalness one for the wider state might also
 allow for an approximate ``probabilistic'' interpretation of the
 results obtained for the different components.

 Thus we see the regularization scheme plays a relevant role in the
 compositeness even with the same number of coupled channels and
 identical kernel potentials. In other words, the so-called compositeness
 used in the present way cannot be taken as a model-independent
 quantity. This is not a surprise, but it  reflects the scheme-dependent
 nature of the field renormalization constant, $\tilde{Z}$. Similar
 conclusions have also been reached in
 Refs.~\cite{Nagahiro:2014mba,Hyodo:2013nka}.

To finish this subsection, we should note that in the present approach, we have only fitted
the mass of the $\Lambda_c(2595)$, while the compositeness
coefficients $\tilde{X}_i$ in Eq.~\eqref{defxi} depend also on the
couplings, which are in turn related to the width.  Note that except in the
naturalness scheme, the predicted width for the $\Lambda_c(2595)$ turns
out to be much larger than its experimental value. A dedicated study
including the isospin breaking effects, together with other channels,
may provide further insight into the problem (see, e.g.,
Ref.~\cite{Guo:2016wpy}), which is however beyond the scope of the
present study.
 
\begin{table}
\begin{center}
\begin{tabular}{c|c|c|c}
  \hline\hline
  coupled channels      & DR-naturalness         & cutoff & DR-HQS  \\ \hline
  Pole position (MeV)    &$2592.25-i0.16$                  & $2592.25-i9.18$         & $2592.25-i3.83$ \\
  $\pi \Sigma_{c}$       & $-0.024+i0.107$       & $0.319+i0.637 $  & $-0.137+i0.960$\\
  $D N$                  & $0.292-i0.026$        & $0.025+i0.018 $  & $0.343-i0.277 $ \\
  $\eta \Lambda_{c}$     & $0.009-i0.001 $       & $0.004-i0.001 $  & $0.040-i0.042 $\\
  $D^{*} N$              & $0.451-i0.055 $       & $0.155-i0.044 $  & $0.243-i0.302 $\\
  $K \Xi_{c}$            & $0.001-i0.000 $       & $0.000-i0.000 $  & $0.001-i0.001 $\\
  $\omega \Lambda_{c}$   & $0.001-i0.000 $       & $-0.000-i0.001$  & $0.014-i0.012 $\\
  $K \Xi^{'}_{c}$        & $0.000+i0.000 $       & $0.000-i0.001 $  & $0.002-i0.001 $\\
  $D_{s} \Lambda$        & $0.026-i0.003 $       & $0.004-i0.000 $  & $0.018-i0.019 $\\
  $D^{*}_{s} \Lambda$    & $0.057-i0.006 $       & $0.008-i0.001 $  & $0.051-i0.054 $\\
  $\rho \Sigma_{c}$      & $0.005-i0.000 $       & $-0.000-i0.002$  & $0.007-i0.004 $\\
  $\eta^{'} \Lambda_{c}$ & $0.018-i0.002 $       & $0.003-i0.000 $  & $0.018-i0.019 $\\
  $\rho \Sigma^{*}_{c}$  & $0.006-i0.001 $       & $0.003-i0.002 $  & $0.006-i0.008 $\\
  $\phi \Lambda_{c}$     & $-0.000-i0.000$       & $-0.000-i0.000$  & $0.000-i0.000 $\\
  $K^{*} \Xi_{c}$        & $0.000+i0.000 $       & $0.000-i0.000 $  & $0.001-i0.001 $\\
  $K^{*} \Xi^{'}_{c}$    & $0.000-i0.000 $       & $-0.000-i0.000$  & $-0.000-i0.000$\\
  $K^{*} \Xi^{*}_{c}$    & $0.000-i0.000 $       & $0.000-i0.000 $  & $0.000-i0.000 $\\
     \hline
$\sum_i \tilde{X}_i     $               & $0.843+i0.012 $       &
     $0.521+i0.602 $  & $0.607+i0.219 $\\
$P$ [Eq.~(\ref{eq:defP})] & 0.001 & 0.565 & 0.095\\
  \hline\hline
\end{tabular}
\caption{Compositeness $\tilde{X}_i$ of each of the 16 coupled channels for the narrow state corresponding to the $\Lambda_c(2595)$.  The potentials $V$ are those of
the SU(6)$\times$HQSS model of Refs.~\cite{GarciaRecio:2008dp,Romanets:2012hm}. The
finite (renormalized) meson-baryon loop function is fitted to the $\Lambda_c(2595)$ mass. This leads to  the following parameters:
$\alpha=0.97952$, $q_\mathrm{max}=0.67898$ GeV, 
$a=-3.37865$ in  the DR-naturalness,  cutoff and the DR-HQS schemes,
respectively.  The real parts of the $\tilde{X}_i$ coefficients, 
calculated within the
DR-naturalness renormalization scheme, were already given 
in Table IV of Ref.~\cite{Garcia-Recio:2015jsa}. According to
Ref.~\cite{Guo:2015daa}, it is only meaningful to give a probabilistic
interpretation to 
$|\tilde{X}_{\pi\Sigma_c}|$.  }\label{Table:narrow}
\end{center}
\end{table}

\begin{table}
\begin{center}
\begin{tabular}{c|c|c|c}
  \hline\hline
  coupled channels      & DR-naturalness     & cutoff  & DR-HQS   \\ \hline
  Pole position (MeV)       &$2606.7-i32.4$                 & $2572.2 $         & $2627.9-i37.4$ \\
  $\pi \Sigma_{c}$       & $0.307+i0.429 $        & $0.041 $   & $0.494+i0.109 $\\
  $D N$                  & $0.005-i0.044 $        & $0.254 $  & $-0.115+i0.001$ \\
  $\eta \Lambda_{c}$     & $0.000+i0.000 $        & $0.009 $   & $0.014+i0.024 $\\
  $D^{*} N$              & $0.048+i0.024 $        & $0.278 $   & $0.322+i0.172 $\\
  $K \Xi_{c}$            & $-0.000+i0.000$        & $0.001 $   & $-0.000+i0.001$\\
  $\omega \Lambda_{c}$   & $0.001-i0.006 $        & $0.001 $   & $-0.005+i0.002$\\
  $K \Xi^{'}_{c}$        & $0.001-i0.005 $        & $0.000 $   & $-0.001-i0.004$\\
  $D_{s} \Lambda$        & $-0.000+i0.001$        & $0.012 $   & $0.006+i0.011 $\\
  $D^{*}_{s} \Lambda$    & $0.001+i0.002 $        & $0.021 $   & $0.016+i0.029 $\\
  $\rho \Sigma_{c}$      & $0.013-i0.027 $        & $0.002 $   & $0.000-i0.012 $\\
  $\eta^{'} \Lambda_{c}$ & $0.000+i0.001 $        & $0.007 $   & $0.007+i0.011 $\\
  $\rho \Sigma^{*}_{c}$  & $0.007-i0.006 $        & $0.002 $   & $0.015+i0.001 $\\
  $\phi \Lambda_{c}$     & $-0.000-i0.000$        & $-0.000$   & $0.000+i0.000 $\\
  $K^{*} \Xi_{c}$        & $0.002-i0.004 $        & $0.000 $   & $0.000-i0.002 $\\
  $K^{*} \Xi^{'}_{c}$    & $0.000-i0.002 $        & $0.000 $   & $0.001-i0.001 $\\
  $K^{*} \Xi^{*}_{c}$    & $0.000-i0.001 $        & $0.000 $   & $-0.000-i0.001$\\
\hline
  $ \sum_i \tilde{X}_i     $               & $0.388+i0.363 $        & $0.616 $   & $0.755+i0.339 $\\
$P$ [Eq.~(\ref{eq:defP})] & 0.243 & 0. 000 & 0.246\\
 \hline \hline
\end{tabular}
\caption{Same as in Table~\ref{Table:narrow}, but for the broader sibling of the
$\Lambda_c(2595)$ resonance.}\label{Table:broad}
 \end{center}
\end{table}

\subsection{Two channels}
In the unitarized chiral approach of Ref.~\cite{Lu:2014ina}, the
$\Lambda_c(2595)$ resonance is dynamically generated from the
coupled--channel interaction between  only the $\pi\Sigma_c$ and $K
\Xi'_c$ meson-baryon pairs.  As
shown in Table~\ref{Table:twochannel}, all three regularization
schemes considered in this work yield
consistent values for the compositeness coefficients, although all with large
imaginary parts and leading to values of the uncertainty parameter $P$ well above
1/2.  Moreover the values for $\tilde{X}_{\pi\Sigma_c}$ listed in
Table~\ref{Table:twochannel} significantly differ from those obtained
in the 16 channel case of Table~\ref{Table:narrow}.  

The cutoff and the DR-HQS subtraction-constant  needed to fit the
$\Lambda_c(2495)$ mass turn out to be rather natural (see the discussion in Sect. II), while  the $\alpha$
parameter in the DR-naturalness scheme deviates appreciably from 1.

Similar conclusions are drawn in the single channel case,
$\pi\Sigma_c$, independently of the value used for the decay
constant. 

\begin{table}
\begin{center}
\begin{tabular}{c|c|c|c}
  \hline\hline
  coupled channels      & DR-naturalness      & cutoff & DR-HQS \\ \hline
  Pole  position (MeV)       & $2592.25-i12.7$              & $2592.25-i15.6$        & $2592.25-i13.5$ \\
  $\pi \Sigma_{c}$       & $0.215+i0.731$        & $0.196+i0.770$  & $0.225+i0.720$\\
  $K \Xi^{'}_{c}$        & $0.003-i0.006$        & $0.001-i0.002$  & $0.003-i0.007$\\
\hline
  $\sum_i \tilde{X}_i    $               & $0.218+i0.725$        &
$0.196+i0.768$  & $0.228+i0.713$\\
 $P$ [Eq.~(\ref{eq:defP})] & 0.823 & 0.904 & 0.799\\
  \hline\hline
\end{tabular}
\caption{Compositeness $\tilde{X}_i$ for the $\Lambda_c(2595)$
  obtained when only the $\pi \Sigma_{c}$ and $K \Xi^{'}_{c}$ channels
  are considered, as in the chiral approach of
  Ref.~\cite{Lu:2014ina}. For all renormalization schemes, the
  coupled--channel 
matrix potential $V$ is taken from this reference
  (note the approaches of
  Refs.~\cite{GarciaRecio:2008dp,Romanets:2012hm,Liang:2014kra}
  provide the same interaction, since it is fixed by SU(3) chiral
  symmetry). The finite (renormalized) meson-baryon loop function is
  fitted to the $\Lambda_c(2595)$ mass. This leads to the following
  parameters: $\alpha=0.8268$, $q_\mathrm{max}=0.7969$ GeV, and
  $a=-5.3768$  in the DR-naturalness, cutoff and DR-HQS schemes,
  respectively.}\label{Table:twochannel}
\end{center}
\end{table}

\subsection{Three channels}

In the local hidden gauge approach of Ref.~\cite{Liang:2014kra}, three
channels are considered, namely $\pi\Sigma_c$, $DN$, and $\eta
\Lambda_c$. Taking the kernel potentials from
Ref.~\cite{Liang:2014kra}, we calculate the compositeness coefficients
$\tilde{X}_i$ using the three regularization schemes introduced in the previous
subsections. Results are shown in Table~\ref{Table:origin}. We can see
that in the DR-naturalness scheme, the $DN$ channel dominates, while
in the DR-HQS method, the $\pi\Sigma_c$ component is the most
significant.  The renormalization method has an important impact on
the compositeness coefficients, despite all renormalization constants
have been adjusted to reproduce the mass of the $\Lambda_c$ resonance. 

We would like to make a further remark here. In the DR-naturalness scheme,
the consideration of the $DN$ channel has led to a value for $\alpha$
quite close to 1, and an uncertainty parameter $P$
[Eq.~(\ref{eq:defP})] very small, enabling for a ``probabilistic''
interpretation. Note that, however, the $P-$values obtained in the other two
renormalization schemes are larger than 1/2, since in both cases the
imaginary parts of $\sum_i\tilde{X}_i$ are much larger than the real ones.
\begin{table}
\begin{center}
\begin{tabular}{c|c|c|c}
  \hline\hline
  coupled channels      & DR-naturalness        & cutoff & DR-HQS  \\ \hline
  Pole position (MeV)  & $2592.25-i0.86$              & $2592.25-i11.4$        & $2592.25-i12.1$ \\
  $\pi \Sigma_{c}$       &$-0.060+i0.483$        & $0.057+i1.002$  & $0.212+i0.729$\\
  $D N$                  & $0.815-i0.390$        & $0.136-i0.355$  & $0.001-i0.001$\\
  $\eta \Lambda_{c}$     & $0.017-i0.008$        & $0.002-i0.006$  &
  $0.001-i0.001$\\
\hline
  $\sum_i \tilde{X}_i $  & $0.772+i0.085$        & $0.195+i0.641$  &
$0.214+i0.727$\\
$P$ [Eq.~(\ref{eq:defP})] & 0.020 & 0.699 & 0.829\\
  \hline\hline
\end{tabular}
\caption{Compositeness $\tilde{X}_i$ for  the $\Lambda_c(2595)$ resonances
  obtained considering three, $\pi \Sigma_{c}$, $DN$ and $\eta
  \Lambda_{c}$, channels as in the  extended hidden gauge approach of
  Ref.~\cite{Liang:2014kra}.  For all renormalization schemes,  the
  coupled--channel matrix potential $V$ is taken from this 
  reference. The
finite (renormalized) meson-baryon loop function is fitted to the $\Lambda_c(2595)$ mass. This leads to  the following parameters:
$\alpha= 0.96048$, $q_\mathrm{max}=0.67535$ GeV, and $a=-5.6365$  in the DR-naturalness,
cutoff and  DR-HQS schemes, respectively. }\label{Table:origin}
\end{center}
\end{table}

\section{large $N_{c}$ evolution}

The $N_{c}$ counting rules for ordinary $qqq$ baryons lead to
scaling laws $\Gamma_{R}\thicksim \mathcal{O}(1)$, $M_{R}\thicksim
\mathcal{O}(N_{c})$ and $\triangle E\equiv M_{R}-M_{B}-m\sim
\mathcal{O}(1)$, with $M_{B}(m)$ the ground-state baryon (meson)
mass,  for the resonance decay width,  mass  and 
excitation energy, respectively~\cite{Witten:1979kh,Goity:2004pw,Cohen:2003fv}.  For an ordinary
$q\bar{q}$ state, its mass, width and decay constant scale as
$\mathcal{O}(1)$, $\mathcal{O}(1/N_c)$ and
$\mathcal{O}(\sqrt{N_{c}})$, respectively.  For dynamically generated
states, the $N_c-$evolution can deviate strongly from such a
scenario~\cite{Pelaez:2003dy, Pelaez:2006nj, Nieves:2009ez,Guo:2011pa,Guo:2015dha}. Compared to the dynamically generated mesons, a study of
dynamically generated baryonic states is complicated because 
baryon flavor representations  change with $N_c$, when the number of
flavors is larger than two~\cite{
  Karl:1985qy,Dulinski:1988yh,Dulinski:1987er}. Such corrections have been
taken into account in the SU(3) chiral study of the $\Lambda(1405)$ in
Refs.~\cite{Roca:2008kr,Hyodo:2007np}, as well as in the study of
negative parity $s$-wave resonances carried out in
\cite{GarciaRecio:2006wb,GarciaRecio:2006xn}, where  a
SU($2N_F$) SF extension of the chiral SU(3) WT interaction
for an arbitrary number of flavors and colors is derived. 
In the present exploratory work on the $\Lambda_c(2595)$, we will 
present $N_c> 3$ results for the chiral two coupled--channel scenario~\cite{Lu:2014ina}, and only in the strict
$N_c\to\infty$ 
limit, in the case of the SU(6)$\times$HQSS model~\cite{GarciaRecio:2008dp,Romanets:2012hm}.

To obtain the large $N_c$ evolution of the
dynamically generated states in  unitarized
approaches, one needs to know how the masses of the
interacting hadrons, the two body loop
function, and the interactions evolve as a function of $N_c$. The latter evolution
is partially a consequence of the change of the flavor representation of the baryons.
In what follows, we examine the $N_c$ dependence of all these inputs.

\subsection{Baryon and meson masses}

Ground-state heavy flavor baryon masses in the $1/m_Q$ and $1/N_c$
expansions have been studied in
Refs.~\cite{Jenkins:1996de,Jenkins:1996rr,Jenkins:2007dm}. Up to
leading order in $1/N_c$, one has 
\begin{equation}\label{baryonmassNc}
  M_{i} = m_Q+M_0 \frac{N_{c}}{3}+ \delta_{i},
\end{equation}
where $m_Q$ is the $N_c$ independent heavy quark mass, $M_0/3$ the
contribution of the light $u$, $d$, $s$ quarks, and $\delta_i$ the
flavor SU(3) breaking contributions. For the present study, we take
$m_Q=m_c=1.275$ GeV, $M_0 \sim 0.9$ GeV, and $\delta_i$ is chosen such
that $M_i$ equals to its physical value for $N_c=3$.  The pseudoscalar meson
masses scale as $\mathcal{O}(1)$ and are taken as constants, while the
pseudoscalar decay constant scales as $\mathcal{O}(\sqrt{N_c})$,
namely,

\begin{equation}
f(N_c)=f_0\sqrt{\frac{N_c}{3}}, \qquad f_0=f(N_c=3).
\end{equation}

\subsection{Loop function}

As already mentioned, the meson-baryon loop function in Eq.~(\ref{eq:loop}) is
logarithmically divergent and   should be regularized. For that
purpose in this work we have used either the dimensional
regularization method or have included a momentum  cutoff to render the
ultraviolet contributions finite. This latter scheme, Eq.~(\ref{eq:cutoff}), is
particularly useful, because its extension to
arbitrary $N_c$ might be more transparent.  

For $N_c=3$, the cutoff takes values of the order of 1 GeV.  Although
the $N_c$ behavior of the cutoff is not known from QCD, it is, however,
clear that within the chiral approach used in Ref.~\cite{Lu:2014ina},
it cannot grow faster than the cutoff of the effective theory itself,
which is of the order of the scale of symmetry breaking $\Lambda_\chi
\sim 4\pi f$.  Otherwise, we would have the absurd situation that we
can extend the validity of the loop integral beyond the applicability
of the theory. Therefore, a natural integral cutoff, as is the case
here, could scale as $\sqrt{N_c}$, but not
faster~\cite{Geng:2008ag}. We will also consider the possibility that
the cutoff may scale slower than $\sqrt{N_c}$, since it would be
$\mathcal{O}(1)$, if it were determined by the existence of heavier
$qqq$ states, which cannot be generated from low-energy baryon-meson
dynamics, and therefore have been integrated out. We will present
results for both scenarios, which yield consistent conclusions, as
it will be shown below.

In the dimensional regularization scheme, the mayor
problem arises from the unknown $N_c$ dependence of the subtraction
constant, $a$. However, in the DR-naturalness scheme, it is given in
terms of the meson and baryon masses~\cite{Hofmann:2005sw,GarciaRecio:2008dp}, which in
turn fix the full dependence of the loop function on $N_c$. This
scheme was employed in Ref.~\cite{GarciaRecio:2006wb} to study the
properties of the negative parity $s$-wave resonances in the large
$N_c$ limit, starting from a SU(6) spin--light flavor extension of
the chiral WT interaction for $N_c=3$. Indeed, some expressions given
in that reference were more general, and can be applied to the
SU($2N_F$) group symmetry for an arbitrary $N_c$. We will take advantage of
these findings and will use the framework set up in
Refs.~\cite{GarciaRecio:2006wb,GarciaRecio:2006xn} to discuss the strict
$N_c\to\infty$  limit of the SU(6)$\times$HQSS
model used in Refs.~\cite{GarciaRecio:2008dp,Romanets:2012hm}.

\subsection{$N_c$ dependence of the meson-baryon interaction}

\subsubsection{$K \Xi'_c-\pi\Sigma_c$ chiral interaction}

In the unitary approach of Ref.~\cite{Lu:2014ina}, the
$\Lambda_{c}(2595)$ resonance is dynamically generated from the chiral
interaction between the pseudoscalar octet of Goldstone bosons and the
sextet ($\Sigma_c,\Xi'_c$) of charmed baryons\footnote{In the heavy
  quark limit, the spin-parity of the light degrees of freedom in
  these baryons is $1^+$.}. In the strangenessless ($S=0)$ isoscalar
($I=0$) sector the interaction reads~\cite{Lu:2014ina}
\begin{eqnarray}
V^{I=0,S=0}(s) = \frac{C^{I=0,S=0}}{4f^2}(E_m+E'_m), \label{eq:defV}
\end{eqnarray}
with $E_m$ and $E'_m$ the center of mass energies of the initial and
final mesons, respectively and the coupled--channel matrix is given by
\begin{eqnarray}
&&\begin{matrix}
      ~~~~K \Xi'_c ~~~~\pi\Sigma_c
\end{matrix}\nonumber \\
 C^{I=0,S=0} &=& \left(
  \begin{matrix}
 -2 & -\sqrt{3} \cr -\sqrt{3} & -4 
\end{matrix}\right) \quad 
\begin{matrix}
      K \Xi'_c \cr \pi\Sigma_c
\end{matrix} ,
\end{eqnarray} 
 In the SU(3) group theory language we have:
\begin{equation}\label{22}
  8\otimes 6 = \overline{3} \oplus 6 \oplus \overline{15} \oplus \overline{24}.
\end{equation}
Although the decomposition involves four SU(3) irreducible
representations, only the $\overline{3}$ and $\overline{15}$ appear in
the $I=0, S=0$ sector. Thus, the coupled--channel matrix  $C^{I=0,S=0}$
becomes diagonal in the
$\left\{|\overline{3}; I=0,S=0\rangle,|\overline{15};
I=0,S=0\rangle\right\}$ SU(3) basis. The meson-baryon and the SU(3)
bases are related by means of an orthogonal matrix $U$ obtained from
the appropriate SU(3) Clebsch-Gordan
coefficients~\cite{GarciaRecio:2010vf}
\begin{equation}
\Big( |\overline{3}\rangle,|\overline{15}\rangle\Big) = \Big( |K \Xi'_c\rangle,|\pi\Sigma_c\rangle\Big)\times U, \qquad
U = \left(
\begin{matrix}
-\frac12 &   -\frac{\sqrt3}{2}\cr -\frac{\sqrt3}{2} & \frac12
\end{matrix}
 \right) .
\end{equation}
In the SU(3) basis, the interaction of Eq.~(\ref{eq:defV}) reads 
\begin{equation}
C^{I=0,S=0}_{\rm SU(3)} = U^\dagger C^{I=0,S=0} U = \left(
\begin{matrix}
-5 &  0 \cr
0 & -1
\end{matrix}
\right).\label{eq:Csu}
\end{equation}
While in the meson sector, the flavor representation remains the
same with the increase of $N_{c}$, the situation in the baryon
sector is more complicated because of the nontrivial variation 
of the flavor representation of the baryons with $N_c$, when the
number of flavors is larger than 2~\cite{Karl:1985qy,Dulinski:1988yh}.  We use the notation $[p, q]$ for an irreducible representation
of SU(3), whose corresponding Young tableau has $p+q$ and $q$ boxes in
the first and second rows, respectively. To extend the irreducible flavor
representation from $N_c=3$ to arbitrary $N_c$, we adopt the
prescription~\footnote{There are two other alternative ways to perform
  the extension. The one  used in the present work,
referred to as the standard one in Ref.~\cite{Dulinski:1988yh},  has
the advantage of keeping the spin, isospin, strangeness and charm
quantum numbers of the original representation at $N_c=3$, while the baryons have different
charge and hypercharge from those at $N_c=3$. }
\begin{equation}\label{33}
  [p,q] \rightarrow [p,q+\frac{N_c-3}{2}],
\end{equation}
For arbitrary $N_{c}$, the 6, $\overline{3}$, and $\overline{15}$
irreducible representations become (we use the notation that an
$N_c-$representation ``$R$'' reduces to $R$ at
$N_c=3$~\cite{Karl:1985qy,Dulinski:1988yh,Dulinski:1987er}),
\begin{equation}\label{su3Nc}
  \begin{split}
    ``6"        & =[2,\frac{N_c-3}{2}], \\
    ``\overline{3}"  & =[0,\frac{N_c-1}{2}], \\
    ``\overline{15}" & =[1,\frac{N_c+1}{2}].
  \end{split}
\end{equation}
From group theory  the SU(3) basis coupling
strengths (eigenvalues)  for arbitrary $N_{c}$ turn out to be (see
Table III of Ref.~\cite{Hyodo:2006kg}):
\begin{equation}\label{CsuNc}
  C^{I=0,S=0}_{\rm SU(3)}(N_{c})= \left(
             \begin{array}{cc}
               -5 &  0 \\
               0  & -\frac{5-N_{c}}{2} \\
             \end{array}
           \right),
\end{equation}
which reduces to Eq.~(\ref{eq:Csu}) at $N_{c}=3$. Note that the
$``\overline{15}"$ eigenvalue becomes repulsive for $N_c> 5$, while
the interaction in the $``\overline{3}"$ subspace is always attractive
and independent of $N_c$, besides the scaling of the decay constant
and masses in Eq.~(\ref{eq:defV}). 

The transformation matrix $U$ will now depend on $N_{c}$ as well. It can be
obtained from the appropriate $N_c$ dependent SU(3) Clebsch-Gordan coefficients. Using
the recursion relations of Ref.~\cite{Cohen:2004ki} or the results of
Ref.~\cite{Hecht:1965NP},  one can easily obtain the explicit form of
$U(N_{c})$ for the decomposition  $8\otimes ``6"= ``\overline{3}"
\oplus ``\overline{15}"\oplus ``6" \oplus``\overline{24}"$.

Following the usual convention, the SU(3) Clebsch-Gordan (CG) coefficients can be expressed as the products of isoscalar factors and ordinary SU(2) CGCs.
\begin{equation}\label{SU3isoscalar}
  \left(
             \begin{array}{cc}
               R_{1}               &  R_{2} \\
               I_{1},I_{1z},Y_{1}  & I_{2},I_{2z},Y_{2} \\
             \end{array}
             \right|
             \left.\begin{array}{c}
             R_{\gamma} \\
             I,I_{z},Y \\
             \end{array}
           \right)=\left(
             \begin{array}{cc}
               R_{1}        &  R_{2} \\
               I_{1},Y_{1}  & I_{2},Y_{2} \\
             \end{array}
             \right|\left.
             \begin{array}{c}
             R_{\gamma} \\
             I,Y \\
             \end{array}
             \right)
             \left(
             \begin{array}{cc}
               I_{1}   &  I_{2} \\
               I_{1z}  & I_{2z} \\
             \end{array}
             \right|\left.
             \begin{array}{c}
             I \\
             I_{z} \\
             \end{array}
             \right)
\end{equation}
where the label $R$ indicates the SU(3) representation, which can be denoted using the usual weight diagram notation $(\lambda,\mu)$, and $\gamma$ labels degenerate representations occurring in a given product.

With the formula given in Table 4 of Ref.~\cite{Hecht:1965NP}, the
transformation matrix $U$ can be obtained  straightforwardly. The
first element, for instance, should be
\begin{equation}\label{U11}
  U_{11}=\sqrt{\frac{(p+1)(\lambda-1-p)q(\lambda+\mu+1-q)(\lambda+\mu+2-q)}{\lambda(\lambda+1)(\mu+1)(\lambda+\mu+2)(\mu+p-q+2)}}
\end{equation}
with 
\begin{equation}\label{U11-2}
  p=\frac{Y}{2}+I+\frac{\lambda'-\mu'}{3},\quad q=\frac{Y}{2}-I+\frac{\lambda'+2\mu'}{3},
\end{equation}
and $ Y$ is related with the $\epsilon$ of
Ref.~\cite{Hecht:1965NP} via $Y=-\epsilon/3$. For the present case, $Y=(N_c-1)/3$ and $I=0$. $(\lambda',\mu'$) refer
to the representation labeled by $``\bar{3}"$ and $``\bar{15}"$ and
their values are given in Eq.~(\ref{su3Nc}).  Keeping in mind that the
formula above is used to calculate the isoscalar factors of
$``6"\otimes 8$, an extra step is needed to obtain the $U$ matrix for
$8\otimes ``6"$. Finally, the $U$ matrix can be written as
\begin{equation}\label{UNc}
  U(N_{c})= \left(
             \begin{array}{cc}
               -\sqrt{\frac{2}{5+N_{c}} }       & -\sqrt{\frac{3+N_{c}}{5+N_{c}}} \\
               -\sqrt{\frac{3+N_{c}}{5+N_{c}}}  & \sqrt{\frac{2}{5+N_{c}}} \\
             \end{array}
           \right).
\end{equation}

With all these ingredients, we finally obtain the $K
\Xi'_c-\pi\Sigma_c$ coupled--channel interaction for an arbitrary
number of colors $N_c$
\begin{equation}\label{eq:CINc}
  C^{I=0,S=0}(N_c)= U(N_c)\left[ C^{I=0,S=0}_{\rm SU(3)}(N_c)\right] U^\dagger(N_c)=\left(
             \begin{array}{cc}
               \frac{N_{c}-7}{2}          & -\sqrt{\frac{N_{c}+3}{2}} \\
               -\sqrt{\frac{N_{c}+3}{2}}  &           -4              \\
             \end{array}
           \right).
\end{equation} 
It is interesting to note that the $\pi\Sigma_c\to \pi\Sigma_c$
interaction is attractive and does not change with $N_c$, while the
$K\Xi'_c$ self-interaction, which is attractive at $N_c=3$, becomes
repulsive for $N_c>7$. On the other hand, the strength of the off-diagonal transition 
increases with $N_c$.

\subsubsection{SU(6)$\times$HQSS}
\label{sec:su8nc}

To better understand the $N_c\gg 1$ limit of the SU(6)$\times$HQSS
model, we need to give some further details on its main features. The
16 coupled--channel model implemented in
Refs.~\cite{GarciaRecio:2008dp,Romanets:2012hm} has its origin in the
compatibility between SF and chiral symmetries, which implies that the
WT interaction can be extended to enjoy SF invariance
[SU(2$N_F)$]. Actually this can be done in a unique way, as it was
demonstrated in ~\cite{GarciaRecio:2005hy}. The model respects SF
symmetry in the light sector and HQSS in the heavy one, and it reduces
to SU(3) WT in the light sector respecting chiral symmetry. HQSS
connects vector and pseudoscalar mesons containing charmed quarks. On
the other hand, chiral symmetry fixes the lowest-order interaction
between Goldstone bosons and other hadrons in a model-independent way;
this is the WT interaction.

As required by SF symmetry, the model of
Refs.~\cite{GarciaRecio:2008dp,Romanets:2012hm} incorporates ground state vector meson and $J^P=3/2^+$
baryon degrees of freedom, in addition to the ground state
pseudoscalar mesons  and $J^P=1/2^+$ baryons. In the large $N_c$
limit, SF becomes exact for the baryon sector~\cite{Dashen:1993jt}. As
for mesons, the lowest-lying states can also be classified quite
naturally according to SF multiplets. Though for charmed mesons SF symmetry
reduces to HQSS, the symmetry works worse for
the light meson spectrum.  

 SF guarantees HQSS except when there are simultaneously $c$ quarks
 and $\bar c$ antiquarks. This is because SF implies invariance under
 equal rotations for $c$ and $\bar c$, but HQSS also requires
 invariance when the two spin rotations are different. Thus, SF does
 not guaranty HQSS in sectors with hidden charm, regardless of
 whether they have net charm or not. As mentioned in
 Subsect.~\ref{sec:16}, in the study of the $C ({\rm charm})=1$ sector
 carried out in Refs.~\cite{GarciaRecio:2008dp,Romanets:2012hm} the WT
 SU(8) interaction kernel was modified, besides using physical masses
 and weak decay constants, by neglecting the hidden charm $cc\bar c$
 channels to accomplish HQSS. The model was quite successful and
 it naturally led to the dynamical generation of the $J^P=1/2^-$
 $\Lambda_c(2595)$ and $J^P=3/2^-$ $\Lambda_c(2625)$ resonances, among
 others. Moreover, it could be used to classify the predicted states
 in SU(6)$\times$HQSS multiplets~\cite{Romanets:2012hm}. Its extension
 to the bottom sector~\cite{GarciaRecio:2012db} easily accommodated
 two narrow baryon resonances with beauty recently observed by the
 LHCb Collaboration~\cite{Aaij:2012da}, that should be intimately
 related to the charmed $\Lambda_c(2595)$ and $\Lambda_c(2625)$ states.

We do not have the mathematical tools to extend the
SU(6)$\times$HQSS model to an arbitrary number of colors, and this is
beyond the scope of this work. However, some results for  the
SU($2N_F$)  WT interaction and an arbitrary number of colors were
obtained in Refs.~\cite{GarciaRecio:2006wb,GarciaRecio:2006xn}. The SU($2N_F$)  WT interaction
for each $JISC$ sector\footnote{Here $J$ stands for the total spin of
  the meson-baryon pair, and for $N_F> 4$, additional flavor quantum
  numbers would need to be specified. }
reads as that in Eq.~(\ref{eq:defV}), but replacing the
coupled--channel matrix there by the appropriate one, $C^{JISC}$, in each sector. Thus for instance, in the
  $\Lambda_c(2595)$ sector,
  the dimension of the coupled--channel space is 21: the sixteen channels
  enumerated in   Tables \ref{Table:narrow} and
 \ref{Table:broad} plus the hidden charm channels, $\Lambda_c\eta_c$,
 $\Lambda_c J/\Psi$, $\Xi_{cc} \bar D$, $\Xi_{cc}\bar D^*$ and
 $\Xi_{cc}^*\bar D^*$. These latter five channels were neglected in
 Refs.~\cite{GarciaRecio:2008dp,Romanets:2012hm} to restore HQSS
 symmetry. As discussed in Refs.~\cite{GarciaRecio:2008dp,Romanets:2012hm} ,
 the SU(8) group reduction\footnote{For any $N_F$, there always
   appears four irreducible representations in the group reduction  of
Eq.~(\ref{eq:su8}). Obviously, the dimensions of them, as well as those
of the representations where ground state baryons and mesons are
included depend on $N_F$. These latter ones are always the adjoint and
the three quark fully symmetric representations, respectively.}
\begin{eqnarray}
\sesentaytres \otimes \cientoveinte = \cientoveinte \oplus
\cientosesentayocho \oplus \dosmilquinientosveinte \oplus
\cuatromilsetecientoscicuentaydos \,,
\label{eq:su8}
\end{eqnarray}
shows that in the SU(8) basis, there exist only four eigenvalues,
associated to each of the irreducible representations that appear on the
right hand side of  Eq.~(\ref{eq:su8}). (Note that the SU(4)
\quince-plet of pseudoscalar ($D_s, D, K, \pi,\eta,\eta_c, {\bar K},
{\bar D}, {\bar D}_s$) and the \dieciseis-plet of vector ($D_s^*,
D^*,K^*, \rho,\omega, J/\Psi, {\bar K}^{*}$, ${\bar D}^*, {\bar D}_s^*,
\phi$) mesons are placed in the $\sesentaytres$ representation.  The
lowest--lying baryons are assigned to
the $\cientoveinte$ of SU(8).  This is appropriate  because in the light sector it can accommodate
 an octet of spin--$1/2$ baryons and a decuplet of
spin--$3/2$ baryons which are precisely the SU(3)--spin
combinations of the low--lying baryon states ($N,\Sigma,\Lambda,
\Xi$ and $\Delta$, $\Sigma^*$, $\Xi^*$, $\Omega$). The remaining
states in the $\veinte_{J=1/2}$ and $\veinte^\prime_{J=3/2}$ are
completed with the charmed baryons: $\Lambda_c$, $\Sigma_c$,
$\Xi_c$, $\Xi'_c$ ,$\Omega_c$, $\Xi_{cc}$, $\Omega_{cc}$ and
$\Sigma^*_c$, $\Xi^*_c$, $\Omega_c^*$, $\Omega^*_{cc}$,
$\Xi^*_{cc}$, $\Omega_{ccc}$, respectively.

The eigenvalues associated to the decomposition of Eq.~(\ref{eq:su8})
were calculated in \cite{GarciaRecio:2006wb,GarciaRecio:2006xn}, for an arbitrary number
of colors and not only for SU(8), but for SU($2N_F$) in
general, and are compiled here in
Table~\ref{tab:groupth}. Independently of $N_c$, in the group
reduction that generalizes Eq.~(\ref{eq:su8}), there
only appear four irreducible representations~\cite{GarciaRecio:2006wb}. For four
flavors, the $\Lambda_c(2595)$ state belongs to the
attractive $\cientosesentayocho$
representation~\cite{Romanets:2012hm}, whose attraction linearly grows
with $N_c$. In this subspace, and keeping in mind the $1/f^2$ factor,
the WT is always attractive and it scales as $\mathcal{O}(1)$, in the
large $N_c$ limit. However in the subspaces associated to the other three representations, the WT
interaction is either repulsive or suppressed,
$\mathcal{O}(1/N_c)$, when $N_c\gg 3$.
\begin{table}
\begin{center}
\begin{tabular}{cc|cc}
\hline\hline
D  & $\lambda_D$ & ``D" & $\lambda_{\rm ``D"}$    \\
\hline
\multicolumn{2}{c|}{$[N_F=4,N_c=3]$} \\
 120& $-16$& $d(N_F,N_c)= \frac{(2N_F+N_c-1)!}{(2N_F-1)! N_c!}$ & $-4N_F$\\
 168& $-22$& $\frac{(2N_F-1)(N_c-1)}{(2N_F+N_c-1)}\times d(N_F,N_c)$ & $-2(N_c+2N_F)$\\
2520& 6& $\frac{(2N_F-1)(2N_F+N_c+1)}{(N_c+1)}\times d(N_F,N_c)$ & $2N_c$\\
4752 & $-2$& $ \frac{2N_FN_c(2N_F+N_c)(2N_F-2)}{(N_c+1)(2N_F+N_c-1)}\times d(N_F,N_c)$ & $-2$\\
\hline\hline
\end{tabular}
\caption{Dimensions (D) and WT eigenvalues ($\lambda_D$) associated to the SU($2N_F$)
  irreducible representations that appear in the  group
  decomposition that generalizes Eq.~(\ref{eq:su8}) [$N_F=4$ and $N_c=3$], for arbitrary number of
  flavors and colors. It corresponds to the reduction of the product
  of the SU(8) adjoint (mesons) and the $N_c-$quark fully symmetric
  (baryons) representations (see Eq.~(27) of Ref.~\cite{GarciaRecio:2006wb}). Note also a misprint  in the
  expression given in Ref.~\cite{GarciaRecio:2006xn} for the dimension of the ``2520" representation.}\label{tab:groupth}
\end{center}
\end{table}

In the SU(8) basis, the coupled--channel interaction matrix $C_{{\rm SU}(2N_F)}^{JISC}$ is diagonal, however
we do not know, for arbitrary $N_c$, the orthogonal matrix
$U_{\rm SU(8)}(N_c)$ that would transform this diagonal matrix into
$C^{JISC}$, the matrix expressed in the meson--baryon basis. It would be
obtained from the appropriate $N_c$ dependent SU(8) Clebsch-Gordan
coefficients\footnote{These coefficients can be found in
  Ref.~\cite{GarciaRecio:2010vf} only for the $N_c=3$ case.}. This
prevents us to obtain the evolution of the $\Lambda_c(2595)$ pole for
moderate values of $N_c>3$, but however as we will discuss in the next subsection, we will be
able to address its behavior for $N_c\gg3$, where we could consider
the loop function diagonal in the meson-baryon basis, as it was done
in Ref.~\cite{GarciaRecio:2006wb}.

\begin{figure}[htpb]
\begin{center}
  \includegraphics[width=0.8\textwidth]{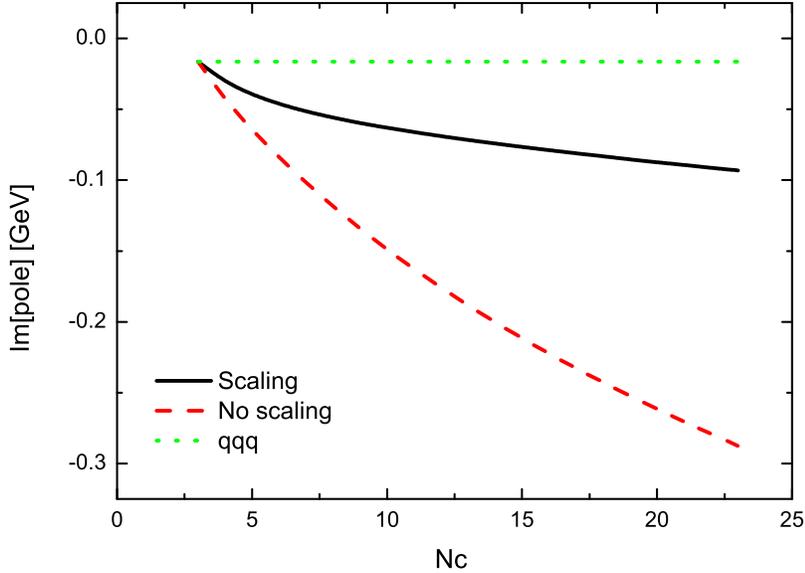}\\
  \caption{Imaginary part of the $\Lambda_c(2595)$ pole position as a
    function of the number of colors. Results have been obtained using
    the $N_c> 3$ extended coupled--channel $K \Xi'_c-\pi\Sigma_c$ chiral interaction
    constructed out Eqs.~(\ref{eq:defV}) and (\ref{eq:CINc}), and
    employing an ultraviolet-cutoff to render the loop function
    finite.  Curves denoted as ``Scaling'' and ``No scaling'' stand
    for the results obtained with different $N_c$ scaling laws for the
    cutoff, either $\mathcal{O}(\sqrt{N_{c}/3})$ or $\mathcal{O}(1)$,
    respectively. }\label{fig:Nc_plot_2ch_Im}
\end{center}
\end{figure}
\begin{figure}[htpb]
 \begin{center}
  \includegraphics[width=0.8\textwidth]{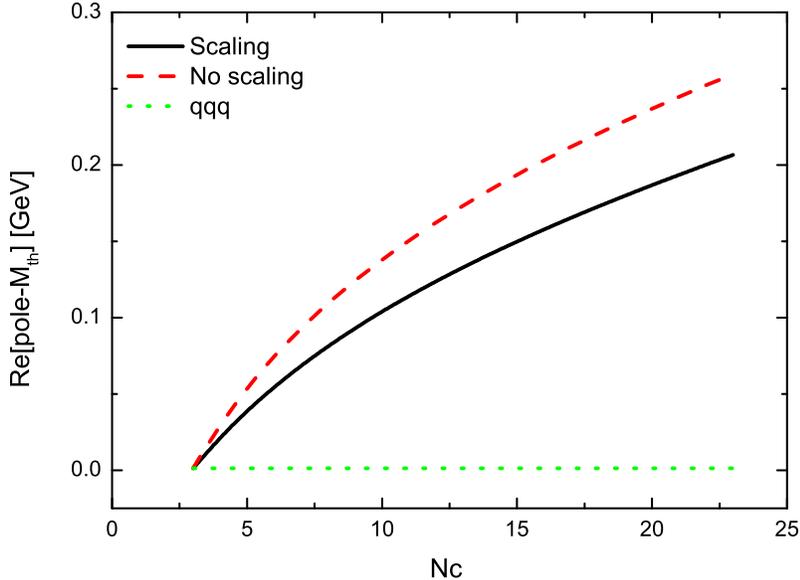}\\
  \caption{The same as in Fig.~\ref{fig:Nc_plot_2ch_Im}, but for the
    real part of the $\Lambda_c(2595)$ pole position, with respect to to the
    $\pi\Sigma_c$ threshold.}\label{fig:Nc_plot_2ch_Re}
\end{center}
\end{figure}

\subsection{$\Lambda_c(2595)$ mass and width for large $N_c$}

 From the findings of the previous subsections it is straightforward to
 study the $N_c$ dependence of the $\Lambda_c(2595)$ mass and width,
 when it is dynamically generated from the coupled--channel 
$K \Xi'_c-\pi\Sigma_c$ chiral interaction.  We use an ultraviolet cutoff to
 renormalize the loop function, and examine two different $N_c$
 scaling laws, $\mathcal{O}(\sqrt{N_{c}/3})$ or $\mathcal{O}(1)$, for
 this parameter of the effective theory.

 Results are displayed in Figs.~\ref{fig:Nc_plot_2ch_Im} and
 \ref{fig:Nc_plot_2ch_Re}, where imaginary and real parts of the
 $\Lambda_c(2595)$ pole position, together with the expected behavior
 of a conventional $qqq$ baryon, are shown as a function of $N_c$. We
 pay attention to moderately large number of colors, up to
 $N_c=25$. For both scaling laws of the cutoff, we find that both mass
 and width of the resonance grow with $N_c$, more rapidly when the
 cutoff is taken as constant. Indeed, the resonance tends to disappear
 since it becomes quite wide (width of hundreds of MeV) and 
 located also hundreds of MeV above the $\pi\Sigma_c$ threshold.  This
 behavior significantly deviates from that expected for a genuine
 $qqq$ state. Thus, the $N_c$ evolution supports the conjecture that
 the meson-baryon component in the wave-function of the
 $\Lambda_c(2595)$ plays a relevant role.

The above analysis is not consistent with the spin symmetry in the
baryon sector, though it only becomes exact in the  large $N_c$
limit~\cite{Dashen:1993jt}, and thus  one should be cautious about
the  consequences extracted in such a scheme.  This has motivated us to
study the $N_c-$evolution of the $\Lambda_c(2595)-$pole position from
a different perspective, implementing exact SU(8) SF symmetry. 

As discussed in Subsect.~\ref{sec:su8nc}, we cannot accurately study
moderate values of $N_c>3$ in this context, because we do not know the
orthogonal matrix $U_{\rm SU(8)}(N_c)$, which implements the change of
basis between the SU(8) one and that constructed out of the meson-baryon
pairs. Yet, even if we knew such rotation, the obtained results for
moderate $N_c$ values would not be physical because SF symmetry does
not guaranty HQSS in this intermediate regime. However, for
sufficiently large values of $N_c$, all meson masses  become
negligible as compared to those of the baryons, all of which in 
turn, to a good approximation, have a common mass $\hat M$, proportional to $N_c$, as inferred from
Eq.~(\ref{baryonmassNc}),
\begin{equation}
\hat M = M_0 \frac{N_c}{3}+\mathcal{O}(1/N_c)
\end{equation}
 In the charm sector $C=1$, there still appear only
two types of configurations involving either only a quark $c$ or an
additional $c\bar c$ pair, since there is always at most only one
charm quark. Since the heavy quark mass is not much larger than
the typical scale associated to the cloud of light degrees of
freedom, and  as $N_c$ increases, the SU(8) SF symmetry should
become more and more accurate. Thus, the pole positions could be obtained in each $JIS$
sector and $C=1$ from (for simplicity, we drop out the label $JISC$)
\begin{equation}
\det\left[I-V(s)G^{II}(s)\right]_{N_c\gg 3}=0
\end{equation}
with $G^{II}(s)$, the matrix loop function calculated in the SRS. In the
DR-naturalness renormalization scheme, $G^{II}(s)$ becomes diagonal in the
meson-baryon coupled--channel basis as it does
the factor $(E_m+E'_m)/f^2 \sim 2(\sqrt{s}-\hat M)/f^2$ in the
definition of the potential\footnote{We are also neglecting SF symmetry
  breaking effects in the weak decay constants.} in
Eq.~(\ref{eq:defV}). Under these circumstances, the resonance position
equation becomes 
\begin{eqnarray}
\det\left[I-V(s)G^{II}(s)\right]_{N_c\gg
  3}&=&\det\left[I-\frac{\sqrt{s}-\hat M}{2f^2}G^{II}(s)U_{\rm
    SU(8)}(N_c)C_{\rm SU(8)}U_{\rm SU(8)}^\dagger(N_c)\right]_{N_c\gg 3}
\nonumber\\
&=&\det\left[U_{\rm
    SU(8)}U_{\rm
    SU(8)}^\dagger-\frac{\sqrt{s}-\hat M}{2f^2}G^{II}(s)U_{\rm
    SU(8)}C_{\rm SU(8)}U_{\rm SU(8)}^\dagger\right]_{N_c\gg 3}
\nonumber\\
&=& \left[\frac{\sqrt{s}-\hat M}{2f^2}G^{II}(s)\right]^n\det\left[\beta(s)-C_{\rm SU(8)}\right]_{N_c\gg 3}=0
\end{eqnarray}
with $\beta(s)= 2f^2/\left[(\sqrt{s}-\hat M)G^{II}(s)\right]$ and
$C_{\rm SU(8)}$ a diagonal matrix constructed out of the four
eigenvalues, $\lambda_{\rm ``D"}$, given in
Table~\ref{tab:groupth}. Besides, $n$ is the dimension of the space
($n=21$ in the $\Lambda_c(2595)$ sector). We see how in the large $N_c$
limit, we can determine the pole position independently of the
orthogonal transformation  $U_{\rm SU(8)}(N_c)$. Thus, the pole positions are determined by 
\begin{equation}
\beta(s)\Big|_{s=s_R\equiv M^2_R-iM_R\Gamma_R} =  \lambda_i, \quad
i=``120", ``168",``2520",``4752"
\label{eq:beta} 
\end{equation} 
with $M_R>M$ and $\Gamma_R>0$. The loop function $G^{II}(s)$ in the
fourth quadrant,
neglecting the meson masses and using a common mass $\hat M$ for the
baryons, can be found in  Eq.~(14) of Ref.~\cite{GarciaRecio:2006wb}. The equation (\ref{eq:beta}) has solutions only for negative
eigenvalues, $\lambda_{``120"}, \lambda_{``168"}$ and
$\lambda_{``4752"}$. As mentioned, the ``168''
irreducible representation of SU(8) leads to the most attractive $s-$wave meson--baryon
interaction, and it becomes the only non-vanishing WT contribution in the
strict  $N_c \to \infty$ limit.

To understand the $N_c$ evolution, the approximated relations of Eqs.~(15), (16)
and~(17) of Ref.~\cite{GarciaRecio:2006wb},
\begin{eqnarray}
\delta^2 \log\delta &\sim& \frac{24\pi^2f_0^2}{N_c\lambda_iM_0^2},
\qquad \delta \equiv \frac{M_R-\hat M}{\hat M}
\label{eq:relsu8a} \\
\frac{\Gamma_R}{M}&\sim & - \frac{\pi \delta}{\log {2\delta}}\sim -\lambda_i \frac{N_c\delta^3M_0^2}{24\pi
  f_0^2}, \quad i= ``120", ``168",``2520",``4752" \label{eq:relsu8b}
\end{eqnarray}
are quite useful. There exist two different situations, neglecting
logarithmic corrections,
\begin{eqnarray}
\lambda_i \sim  \mathcal{O}(1) &\Rightarrow&  (M_R-\hat M)\sim 
\sqrt{N_c}, \quad \Gamma_R \sim \sqrt{N_c} \\
\lambda_i \sim  \mathcal{O}(N_c) &\Rightarrow&  (M_R-\hat M)\sim 
\mathcal{O}(1), \quad \Gamma_R \sim  \mathcal{O}(1)
\end{eqnarray}
From the results of Table III of Ref.~\cite{Romanets:2012hm},
we can see that the two $\Lambda_c(2595)$ states predicted in
Ref.~\cite{Garcia-Recio:2013gaa} and the $J^P=3/2^-$ $\Lambda_c(2625)$
resonance stem
from the $\cientosesentayocho$ representation, and thus one deduces
that their widths and
excitation energies behave as $\mathcal{O}(1)$ for $N_c\gg 3$, as
predicted by Witten almost 30 years ago for genuine $qqq$
states. However, the width and excitation energy of the fourth resonance in
the table, located around\footnote{This resonance, with large
couplings to $\Lambda_c\eta$ and $\Xi_c K$, is also found in
Ref.~\cite{Lu:2014ina}.} 2800 MeV and associated to the $\cientoveinte$
representation, grow as $\sqrt{N_c}$ in this
limit. That is, this resonance would disappear, since it becomes wider
and heavier as $N_c$ increases. This behavior would be similar to 
what we have seen earlier in Figs.~\ref{fig:Nc_plot_2ch_Im} and
\ref{fig:Nc_plot_2ch_Re}. Note that the large
$\cuatromilsetecientoscicuentaydos$ is attractive and contains many
exotic states that would disappear in the large $N_c$ limit as deduced
from the above discussion. 

The fact that the $\Lambda_c(2595)$ resonance survives in the large
$N_c$ limit, contradicting the findings of
Figs.~\ref{fig:Nc_plot_2ch_Im} and \ref{fig:Nc_plot_2ch_Re}, is
however quite natural. Indeed, it is natural to admit the existence of
a (perhaps) sub-dominant $qqq$ component in the resonance wave
function. Indeed, this resonance has been studied with some success using
a constituent quark model in Ref.~\cite{Yoshida:2015tia}. Thus, one might expect the $N_c$ behavior close to the
physical value $N_c=3$ of the resonance is non $qqq$ due to the
unitarity logs, but this sub-dominant $qqq$ component would become dominant when the
number of colors gets sufficiently large~\cite{Nieves:2009ez,Nieves:2011gb}. 

It is interesting to note that recently lattice QCD simulations have
started to probe the dependence on $N_c$ of the properties of 
mesonic~\cite{Bali:2013kia} and baryonic~\cite{DeGrand:2012hd,DeGrand:2013nna}
states. (See,
Ref.~\cite{Lucini:2012gg} for a comprehensive review.) Testing the
$N_c$ dependence of the $\Lambda_c(2595)$ and other proposed molecular
states can help to unravel their true nature. In this sense, the
present study should serve a motivation for such studies.

\section{Summary}
Understanding the Fock components of a hadronic state is a nontrivial
task due to the non-perturbative nature of the strong interactions at
the relevant scales. Recent experimental observation of the so-called
$XYZ$ and baryonic pentaquark states have challenged the conventional
wisdom that baryons are composed of three quarks and mesons of a
quark-antiquark pair. More surprisingly, large hadron-hadron
components are predicted for certain well established hadrons, e.g.,
the $N(1535)$. In the present work, we have used two widely accepted
approaches to qualify the $\Lambda_c(2595)$ as a dynamically generated
state, namely, the compositeness condition and the large $N_c$
evolution. Our results show that, although the relative importance of
a particular coupled channel cannot be determined in a model
independent manner, the basic picture that the $\Lambda_c(2595)$ has
relevant meson-baryon components emerges as a robust conclusion. We
have also shown that the commonly defined compositeness of the state
depends on the included coupled channels, and also on the scheme
adopted to renormalize the ultraviolet divergent meson-baryon loop
function, which appears in the unitarized approaches. The importance of
the molecular picture is also corroborated by our study of the
dependence on the number of colors of the mass and width of the
$\Lambda_c(2595)$. It is shown that for moderate $N_c> 3$ values, they differ
largely from those expected for a genuine $qqq$ state. We can not
however discard the existence of
a (perhaps) sub-dominant $qqq$ component in the resonance wave
function,  which would become dominant when the
number of colors gets sufficiently large.
\begin{acknowledgments}
This work is partly supported by the National Natural Science
Foundation of China (NSFC) under Grant Nos. 1375024, 11522539, 11575052 and 11105038,  by the Spanish Ministerio de Econom\'\i a y
 Competitividad and European FEDER funds under the contracts
 FIS2014-51948-C2-1-P,  FIS2014-57026-REDT and SEV-2014-0398, and by
 Generalitat Valenciana under contract PROMETEOII/2014/0068, the Natural Science Foundation of Hebei Province with contract No.~A2015205205,
the grants from the Education Department of Hebei Province under contract No.~YQ2014034,
the grants from the Department of Human Resources and Social Security of Hebei Province with contract No.~C201400323, 
the Sino-German Collaborative Research Center ``Symmetries and the Emergence of Structure in QCD'' (CRC~110) 
co-funded by the DFG and the NSFC. 
\end{acknowledgments}

\end{document}